\documentclass[pra,twocolumn,superscriptaddress,showpacs,floatfix,nofootinbib]{revtex4}
\usepackage{epsfig,amsmath,amsfonts,amssymb}
\usepackage{graphicx}
\usepackage[utf8]{inputenc}
\usepackage{boxedminipage}
\usepackage{epsfig,psfrag}
\usepackage{amsmath}
\usepackage{mathbbol, amsfonts}
\usepackage{lscape}
\usepackage{fancybox}
\usepackage{amsfonts,amssymb}
\usepackage{color}
\newcommand{\beq}{\begin{equation}}
\newcommand{\eeq}{\end{equation}}

\newcommand{\beqa}{\begin{eqnarray}}
\newcommand{\eeqa}{\end{eqnarray}}
\def\ra{\rangle}
\def\la{\langle}

\usepackage{wrapfig}

\newcommand{\be}{\begin{equation}}
\newcommand{\ee}{\end{equation}}

\begin{document}
\title{Manufacturing time operators: covariance, selection
  criteria, and examples}
\author{G. C. Hegerfeldt}
\affiliation{Institut f\"ur Theoretische Physik, Universit\"at G\"ottingen,
Friedrich-Hund-Platz 1, 37077 G\"ottingen, Germany}
\author{J. G. Muga}
\affiliation{Departamento de Qu\'{\i}mica F\'{\i}sica, Universidad del
Pa\'{\i}s Vasco, Apartado 644, 48080 Bilbao, Spain}
\author{J. Mu\~noz} 
\affiliation{Departamento de Qu\'{\i}mica F\'{\i}sica, Universidad del
Pa\'{\i}s Vasco, Apartado 644, 48080 Bilbao, Spain}
\begin{abstract}
We provide the most general forms of covariant and normalized 
time operators and their probability densities, with  
applications to quantum clocks, the time of arrival,
and Lyapunov quantum operators.  
Examples are discussed of the profusion of possible operators and their 
physical meaning. Criteria to define unique, optimal operators 
for specific cases are given.  
\end{abstract}
\pacs{03.65.Ta,03.65.Ca,03.65.Nk}
\maketitle
\section{Introduction}
Quantum theory provides, for a given state preparation,  
expectation values and distributions for a number of observables
whose operators have been identified by a combination of heuristic
arguments (e.g. quantization rules),
and consistency arguments. Their relevance and validity is eventually
put to the test or motivated by experiments. Time observables, i.e.,
random variables  
such as the arrival times of particles at a detector for a given state
preparation, have been more problematic than other observables like
``energy'', 
``momentum'' or ``position'' evaluated at a fixed instant. In fact 
almost a century after the creation of the basic quantum formalism, the 
theoretical framework to deal with time observables which have a relatively 
straightforward operational definition in the laboratory, 
is still being debated.      
Reviews of several aspects of the difficulties and 
efforts to formalize time in quantum mechanics may be found in two recent
books \cite{b1,b2}.    
Some of these difficulties may be traced back to a lack of a general  
framework to generate and define ``time operators''.
An important point, frequently overlooked, is that, for a given system, 
there is no single time operator. There are infinitely 
many time operators corresponding to different observables and apparatus.
``Canonical time operators'' have been defined \cite{Holevo,Hall},
but, as we shall stress, the definition of ``canonical'' is basis
dependent, even without energy degeneracy. Thus, further analysis is
necessary  
to set ideal operators and possibly uniqueness in some cases 
by imposing the physical conditions to be satisfied (e.g. symmetries)
or optimal properties, such as a minimal variance.

Time operators can be classified into two main groups on physical grounds, 
depending on their association with time durations or time
instants. An example of a duration is the dwell time of a particle in
a region of space.  
The corresponding operator commutes with the Hamiltonian since the
duration of a future process does not depend on the instant  
that we choose to predict it \cite{dwell}. Instead, the other group of
time observables are shifted by the same amount as the preparation
time, either forward 
(clocks) \cite{clocks} or backward (event times recorded with a
stopwatch, the simplest case being the time of arrival), 
and are conjugate to the Hamiltonian.
We shall set here a framework for these ``covariant'' observables \cite{Holevo}
associated with instants and analyze their multiplicity and physical 
properties. Applications are discussed 
for quantum clocks and the time of arrival. The relation to Lyapunov operators
is also spelled out. 

The plan of the paper is as follows.
After introducing the main concepts and notation in Sec. \ref{notation},   
in Section \ref{general} the most general form of a covariant time
operator is determined for a Hamiltonian with only continuous,
possibly degenerate, eigenvalues. In Section \ref{uniqueness} it  
is shown that, for a time-reversal invariant Hamiltonian, one arrives
at a unique and natural form of time operator by imposing time
reversal covariance, invariance under 
additional symmetries and minimality of the variance. 
In Section \ref{arrival} the results are applied to arrival times for
a particle moving on a half-line and a
connection with the delay time of Smith \cite{Smith} is established. 
In Section \ref{Lyapunov} 
the results are applied to Lyapunov operators which were considered in
Ref. \cite{Strauss}. It is shown that the expression given there is
a special case, and the general form as well as possible uniqueness
conditions are presented. In particular it is shown that for a time
reversal invariant Hamiltonian there is no time reversal invariant
Lyapunov operator. This is of interest because it has been argued
that, in order to characterize a quantum system as
irreversible and an arrow of time if the Hamiltonian is time reversal
invariant and if one uses a formulation in terms of Lyapunov
operators, a Lyapunov operator or functional 
should be time reversal invariant \cite{Sewell}.
\section{Covariance of time operators. Notation.}\label{notation}
We differentiate between clock time operators and event time
operators. The former, denoted by $\hat{T}$, can be associated with
a quantum clock which measures the progressing parametric time,
while the latter, denoted by $\hat{T}^A$, describe the time of an event,
for example, the instant of time a particle is found to arrive at a particular position. This and the following two sections are mostly devoted 
to clock observables, although the formal results are analogous for event times.
In an ordinary clock the dial position is the observable which tells us what time it is. In a quantum clock the dial ``position'' is probabilistic but its 
average should follow faithfully the advancement of parametric time.
We would like as well to minimize the variance and estimate the time 
as accurately as possible with a finite number of measurements.    
We will not investigate here specific operational realizations, 
see a review in \cite{clocks}, but instead idealized operators and
their properties.    
\subsection{Clock time operators}
For a given state $|\psi \rangle$, let the probability of finding the
measured time in the interval $(-\infty, \tau)$ be given by the
expectation with $|\psi \rangle$ of an operator $\hat{F}_\tau$. Note
that $0\leq \hat{F}_\tau \leq 1$ so that $\hat{F}_\tau$ is
selfadjoint and bounded. [For a momentum 
measurement the analogous operator would be $\int_{- \infty}^p
dp^\prime |p^\prime \rangle \langle p^\prime|$ for finding the
momentum in $(- \infty, p)$. Here $\hat{F}_\tau$ can have a more
general form and in general one deals with a positive-operator valued
measure.] Then  
\begin{equation}
\label{eq1.1} 
\Pi (\tau; \psi) \equiv \frac{d}{d\tau} \langle \psi | \hat{F}_\tau|
\psi \rangle 
\end{equation}
is the corresponding temporal probability density, normalized as $\int
d\tau \,\Pi(\tau; \psi)= 1$. We define the probability density
operator $\hat{\Pi}_\tau $ by 
\begin{equation}
\label{eq1.2}
\hat{\Pi}_\tau \equiv \frac{d}{d\tau} \hat{F}_\tau,
\end{equation}
normalized as  
\begin{equation}
\label{eq1.2a}
\int_{- \infty}^\infty d\tau\, \hat{\Pi}_\tau = \Eins.
\end{equation}
%
The mean value of observed time can be written as
%
\beq
\int_{- \infty}^\infty\!\! d\tau\, \tau \,\Pi (\tau; \psi) = \langle \psi| \int_{-
\infty}^\infty\!\! d\tau\, \tau\, \hat{\Pi}_\tau | \psi \rangle
\equiv \langle \psi | \hat{T} | \psi \rangle,
\label{eq1.3}
\eeq
%
where
\begin{equation}
\label{eq1.4}
\hat{T} \equiv \int_{- \infty}^\infty d\tau\, \tau\, \hat{\Pi}_\tau
\end{equation}
is called the time operator associated with $\hat{\Pi}_\tau$.
The second moment, if it exists, is given by
\begin{equation}
\label{eq1.5}
\int d\tau ~\tau^2\, \Pi (\tau; \psi) = \langle \psi| \int
d\tau\,\tau^2\, \hat{\Pi}_\tau\,  
|\psi \rangle
\end{equation}
and similarly for higher moments. It may happen that this is not equal to
$\langle \psi | \hat{T}^2| \psi \rangle$.

A clock time operator is called covariant with respect to ordinary
(parametric) time if for the states $|\psi \rangle \equiv | \psi_0
\rangle$ and $|\psi_{t} \rangle$ the probabilities of finding the
measured time in the  respective intervals $(- \infty, \tau)$ and
$(- \infty, \tau + t)$  coincide, i.e. if
\begin{equation}
\label{eq1.6}
\langle \psi_0 | \hat{F}_\tau|\psi_0 \rangle = \langle \psi_{t} |
\hat{F}_{\tau + t}|\psi_{t} \rangle.
\end{equation}
This implies, choosing $t = - \tau$,
\begin{eqnarray}
\label{eq1.7a}
\hat{F}_\tau &=& e^{-i \hat{H} \tau/\hbar} \hat{F}_0\, e^{i \hat{H}\tau/\hbar}\\
\label{eq1.7bc}
\hat{\Pi}_0 &=&  \frac{-i}{\hbar}[\hat H, \hat F_0]\\
\label{eq1.7b}
\hat{\Pi}_\tau &=&
e^{-i \hat{H} \tau/\hbar}\,\hat{\Pi}_0\, e^{i \hat{H}\tau/\hbar}.
\end{eqnarray}
Note that $\la\psi|\hat{\Pi}_\tau|\psi\ra$ is non-negative because
$\la\psi|\hat{F}_\tau|\psi\ra$ is non-decreasing. 
By a change of variable in Eq. (\ref{eq1.4}) one obtains
\begin{equation}
\label{eq1.7c}
e^{i \hat{H}t/\hbar}\, \hat{T} \,e^{-i \hat{H}t/\hbar} = \hat{T} + t.
\end{equation}
{}From this it follows by differentiation that $\hat H$ and $\hat T$
satisfy the canonical commutation relation
\beq
\label{CCR}
[\hat T,\hat H] = i\hbar
\eeq
when sandwiched between (normalizable) vectors from the domain of $\hat H$.

Note that
$\hat \Pi_0$ and $\hat \Pi_\tau$ are in general not operators on the
Hilbert space but  only bilinear forms evaluated
between normalizable vectors from the domain of $\hat H$.  
An expression like $\la E| \hat{\Pi}_0|E'\ra$ has to be understood as a
distribution. Since the diagonal $E=E'$ has measure 0 it is no contradiction
that Eq. (\ref{eq1.7bc}) gives 0 on 
the diagonal while the following example gives $(2\pi\hbar)^{-1}$.

\noindent
{\bf Example}: For a Hamiltonian $\hat{H}$ with non-degenerate
continuous eigenvalues $E$ and
eigenvectors $|E\rangle$ with $\langle E| E^\prime \rangle = \delta
(E - E^\prime)$ we put
\begin{equation}
\label{eq1.13}
\langle E|\hat{\Pi}_0|E^\prime \rangle \equiv \frac{1}{2\pi\hbar},
\end{equation}
so that in this case
\begin{eqnarray}
\label{eq1.14}
\hat{\Pi}_0&=&\frac{1}{2\pi\hbar} \int dE~dE^\prime |E\rangle \langle E^\prime|,
\\ \label{eq1.15}
\hat{\Pi}_\tau &=& \frac{1}{2\pi\hbar}
\int dE~dE^\prime e^{-i(E-E^\prime)\tau/\hbar}
|E\rangle \langle E^\prime|.
\end{eqnarray}
%
The normalization condition of Eq. (\ref{eq1.2a}) is easily checked. The
corresponding clock time operator $-i\hbar\partial_E$
which results from Eq.~(\ref{eq1.4}), has been
considered the ``canonical time operator in the energy
representation'' \cite{Holevo,Hall,Hall08},   
but note that $|E \rangle$ is unique only up to a phase
\cite{Holevo,Caves,Hall95}, 
and taking $|E \rangle_\varphi \equiv e^{i \varphi (E)}| E \rangle$
instead of $|E \rangle$ leads, for different $\varphi$, 
to multiple ``energy representations'',
even for a system without any degeneracy. In the new basis  the
``canonical operator'' will be shifted by   
\begin{equation}
\label{eq1.16}
\hbar\int dE ~\varphi^\prime (E) |E\ra \langle E|.
\end{equation}
Moreover, the mean-square deviation $\Delta T^2$ for a given state
depends on $\varphi(E)$ in such a way that there is no choice of
$\varphi(E)$ which would make $\Delta T$ minimal for {\it all} states,
as shown in Appendix \ref{B}. Therefore, in this case a minimality
condition imposed on $\Delta T$ cannot be fulfilled and does not lead
to a unique natural choice of time operator without further additional
restrictions. There must be additional physical criteria to choose,
and in fact several of them may be physically significant. This will be
exemplified below, see Sect. \ref{arrival}.   
\subsection{Arrival time operators}
``Time-of-event'', and in particular time-of-arrival operators and
probability densities are similar to clock operators (for reviews of
this concept see \cite{MugaLeavens,toatqm2}).  
Physically, we expect that a free particle in one dimension will arrive with certainty at a given detection point (including negative times and ignoring the 
case of zero momentum which is of measure zero for an arbitrary physical wave-packet). Similarly a free particle in three dimensions will arrive at an
infinitely extended plane.   
Also, a particle on a half-line with reflecting
boundary conditions and without additional potential, is expected
to arrive once at the boundary and, at least on classical grounds, 
twice at any other point. In
the latter case it is meaningful to consider the first arrival at a
given point because this should be in principle observable. In all these
cases the total arrival probability resp. first-arrival probability is 1.
The corresponding arrival time
operators are denoted by $\hat{T}^A$ and $\hat{\Pi}^A_t$,
respectively,  and when compared to clock operators their formal properties are 
identical up to a change of sign, e.g. in the conjugacy relations or the 
formulation of covariance \cite{Werner}.  
This means that, in contrast to clock times, if the particle's state is
shifted in time by $t_0$, it 
should arrive a time $t_0$ earlier, and the temporal probability density
should be shifted by $t_0$ to earlier times.    
These are, in other words, waiting times until an event occurs, which depend 
on the time when we set the stopwatch to zero, and decrease if we reset it  
at a later instant. 
Thus the analog of the cumulative probability operator in
Eq. (\ref{eq1.6}) must now satisfy  
\begin{equation}
\begin{split}
\langle \psi_0 | \hat{F}^A_\tau|\psi_0 \rangle &= \langle \psi_{t} |
\hat{F}^A_{\tau - t}|\psi_{t} \rangle\\
\label{eq5.7a}
\hat{F}^A_\tau &= e^{i \hat{H} \tau/\hbar} \hat{F}^A_0\, e^{-i \hat{H}\tau/\hbar}.
\end{split}
\end{equation}
With $\hat{\Pi}^A_t\equiv d\hat{F}^A_t/dt$ and
$\hat{\Pi}^A_0=\frac{i}{\hbar}[\hat{H},\hat{F}^A_0]$  
we have 
\beqa \label{eq5.8a}
\hat{T}^A &=& \int {dt}\, t \, e^{i \hat{H}t/\hbar}\, \hat{\Pi}^A_0 \,
e^{-i \hat{H}t/\hbar}, 
\\
\label{eq5.9a}
\hat{\Pi}^A_{t} &=& e^{i \hat Ht/\hbar}\, \hat{\Pi}^A_0 \, e^{-i\hat Ht/\hbar}, 
\\
%
\label{eq5.10a}
\langle \psi_{t_0} |\hat{T}^A | \psi_{t_0} \rangle &=& \langle \psi_0
|\hat{T}^A| \psi_0 \rangle - t_0,
\\
\nonumber
\langle \psi_{t_0} |\hat{\Pi}^A_t | \psi_{t_0} \rangle &=& \langle \psi_0
|\hat{\Pi}^A _{t + t_0} |\psi_0 \rangle. 
\end{eqnarray}
In addition, the operator should incorporate the location where
the arrivals are observed. For 
free particles  coming in from one side and arrivals at a plane this
was achieved  in Ref.  \cite{Kij} by
postulating invariance of the probability density under a combination of
space reflection and time reversal. 
It is evident that these properties still do not specify the
operator uniquely. For physical reasons one will also demand for an
optimal arrival-time observable that the arrival-time probability density has
minimal variance, analogous to the postulate in Ref. \cite{Kij} for 
free particles in three-dimensional space. This means
that no other arrival-time observable can be measured more precisely. 
\section{The general form of covariant time operators} 
\label{general}  
We begin with covariant clock time operators associated with a given
Hamiltonian $H$. For simplicity, we  first consider the case when
$\hat H$ has only non-degenerate continuous eigenvalues $E$, with
generalized  eigenvector $|E\rangle$ and normalization
\[\langle E |E^\prime \rangle = \delta (E - E^\prime).\]
We will determine the most general form of
$\hat{\Pi}_0$ which, through Eqs. (\ref{eq1.1} - \ref{eq1.7b}), leads to a
covariant probability density operator and corresponding time
operator.

The simple example in Eq. (\ref{eq1.14}) can be generalized to
\begin{equation*}
\hat{\Pi}_0= \frac{1}{2\pi\hbar}\int dE\, dE^\prime\, b(E)\,|E \rangle \langle E^\prime|\,
\overline{b(E^\prime)} 
\end{equation*}
and, more generally, it will be shown that 
\begin{eqnarray}\label{eq2.1}
\hat{\Pi}_0& =& \frac{1}{2\pi\hbar}\sum_i \int dE \,dE^\prime
\,b_i(E)\,|E\rangle \langle E^\prime|\, 
\overline{b_i(E^\prime)},
\\
\hat{T} &=& \frac{1}{2\pi\hbar}\sum_i \int {dt}\, t \int dE\,dE^\prime~
e^{-i(E-E^\prime)t/\hbar}
\nonumber\\
&\times&b_i(E)\,|E\rangle \langle E^\prime|\,\overline{b_i(E^\prime)}\label{tgen}
\end{eqnarray}
is the most general form of $\hat{\Pi}_0$ and $\hat T$, 
where the functions $b_i(E)$  have to satisfy certain properties in
order that the total probability is 1 and that the second moment in
Eq. (\ref{eq1.5})  is finite. Indeed, for given state
$|\psi \rangle$, the total temporal probability is, with $\psi(E)\equiv \langle
E|\psi \rangle$, 
%
\beqa
&&\int_{- \infty}^{+ \infty} dt 
 \langle \psi|\, e^{-i\hat Ht/\hbar}\,
\hat{\Pi}_0\, e^{i\hat H t/\hbar}\,|\psi \rangle 
\\ \nonumber
&&= \sum_i \int \frac{dt}{2 \pi\hbar} \left|\int dE\, e^{-iEt/\hbar}\,
\overline{\psi(E)}\, b_i(E)\right|^2
\\ \nonumber 
&&= \sum_i  \int dE\, dE^\prime \,
\delta(E-E^\prime)\, \overline{\psi (E)}\, b_i (E)\,
\overline{b_i(E^\prime)}\, \psi(E^\prime)
\\ \nonumber
&&=
\sum_i \int dE~\overline{\psi (E)} \sum_i b_i(E)\, \overline{b_i(E)}\,
\psi(E).
\eeqa
This equals 1 for every state $|\psi \rangle$ if and only if
\begin{equation} \label{eq2.3}
\sum_i b_i (E) \,\overline{b_i(E)} = 1.
\end{equation}
Similarly,
\beqa
\!\!\!\!\!\!\langle \psi |\hat{T}| \psi \rangle &=&\int dE\, \bar{\psi} (E)
\,\frac{\hbar}{i}\, \psi^\prime (E) 
\nonumber\\
&&+ \int dE\, |\psi (E)|^2\, \frac{\hbar}{i} \sum b_i(E)\,
\overline{b_i^\prime (E)}.
\label{eq2.3a}
\eeqa
Note that $\sum b_i\, \bar{b_i^\prime}$ is purely imaginary, from
Eq. (\ref{eq2.3}), and thus vanishes if $b_i$ is real. 

The second
moment is 
%
\beqa\label{eq2.4}
&&\int  dt\, t^2
\langle \psi |e^{-i\hat Ht/\hbar}\, \hat{\Pi}_0 \,e^{i\hat Ht/\hbar}|\psi \rangle
\\\nonumber
&&= \hbar\sum_i \int \frac{dt}{2 \pi} \left|\int
dE\, \partial_E\, e^{-iEt/\hbar}\,
\overline{\psi(E)}\, b_i(E)\right|^2
\\\nonumber 
&& =\hbar^2\sum_i \int dE \,\partial_E \left(
\overline{\psi (E)}\, \,b_i(E)\,
\right) \partial_E \left( \overline{b_i(E)}\, \psi (E)  \right)
\\\nonumber
&&= \hbar^2\int dE \left\{
|\psi^\prime(E)|^2 + \sum_i |b_i^\prime (E)|^2 \, |\psi(E)|^2\right. 
\\\nonumber
&&\left.+ 2\,
{\rm Re}\,
\sum_i \overline{b_i(E)}\, b_i^\prime(E)\, \overline{\psi(E)}\, \psi^\prime
(E)\right\}
\eeqa
%
by Eq. (\ref{eq2.3}). This is finite if and only if the contribution
from the first and second term are finite, and for the latter to hold for all
infinitely differentiable functions $\psi(E)$ vanishing outside a finite interval (i.e. with compact support in
$E$) one must have 
\begin{equation} \label{eq2.6}
\sum_i |b_i^\prime(E)|^2  ~~ \mbox{integrable over any finite interval.}
\end{equation}
Eq. (\ref{eq2.1})
gives the most general form of $\hat{\Pi}_0$ leading to a covariant
time operator 
when the functions $b_i$  satisfy Eqs. (\ref{eq2.3}), and the second
moment is finite for states with $\langle E|\psi \rangle$ of compact
support if and only if  Eq. (\ref{eq2.6}) holds. 

For a given
$\hat{\Pi}_0$ one can construct  the functions $b_i$ as follows. One
chooses a maximal set $\{|g_i\rangle\}$ of vectors satisfying
\be \label{eq2.6a}
\langle g_i |\hat{\Pi}_0| g_j \rangle = \delta_{ij}/2\pi\hbar.
\ee
Such a maximal set is easily constructed by the standard Schmidt
orthogonalization procedure. Then a possible set $\{b_i\}$ is given by
\be \label{eq2.6b}
b_i(E) =2\pi\hbar\, \langle E |\hat{\Pi}_0| g_i \rangle.
\ee
Eq. (\ref{eq2.1}) is then a realization of the given $\hat{\Pi}_0$.
Mathematical details, in particular regularity properties, will be
presented elsewhere \cite{Gerhard}. It should be noted 
that the functions $b_i$ in the decomposition of $\hat{\Pi}_0$ in
Eq. (\ref{eq2.1}) are not unique. 

For the case of degenerate eigenvalues of $\hat{H}$ we first consider
the case where the degeneracy is indexed by a discrete number and such that
\begin{equation} \label{eq2.7}
\langle E, \alpha | E^\prime, \alpha^\prime \rangle = \delta_{\alpha
\alpha^\prime}  \delta(E- E^\prime).
\end{equation}
For simplicity we assume the same degeneracy for each $E$. Then
Eqs. (\ref{eq2.1} - \ref{eq2.6}) generalize as
\begin{equation} \label{eq2.8}
\begin{split}
\hat{\Pi}_0 = \frac{1}{2\pi\hbar}\sum_i&\int  dE\,dE^\prime \\
&\sum_{\alpha \alpha'} 
b_i(E, \alpha)\,|E, \alpha \rangle \langle
E^\prime , \alpha^\prime|\, \overline{b_i (E^\prime, \alpha^\prime)},
\end{split}
\end{equation}
\begin{equation} \label{eq2.9}
\sum_i b_i(E, \alpha)\, \overline{b_i (E, \alpha^\prime)}=
\delta_{\alpha \alpha^\prime},
\end{equation}
\begin{equation} \label{eq2.10}
\mbox{second moment} = \hbar^2\int dE\,\Big|\partial_E
\sum_\alpha \overline{b_i (E, \alpha)}\, \psi (E, \alpha) \Big|^2,
\end{equation}
\begin{equation} \label{eq2.11}
\sum_i |b_i^\prime (E, \alpha)|^2~~ \mbox{integrable over any finite
interval} 
\end{equation}
for each $\alpha$, where $\psi (E, \alpha)\equiv \langle E ,
\alpha|\psi \rangle$ and 
$b_i(E,\alpha) = 2\pi\hbar\langle E,\alpha |\hat{\Pi}_0| g_i \rangle $.
Again Eq. (\ref{eq2.8})
gives the most general form of $\hat{\Pi}_0$ leading to a covariant time
operator through Eqs. (\ref{eq1.1}-\ref{eq1.7b}).
The case of continuous degeneracy parameter can be reduced to the
discrete case.

These results carry over in a corresponding way to  arrival times with
normalized probability densities. 
\section{Uniqueness of time operator: time reversal, symmetries  and  minimal variance}\label{uniqueness} 
As seen in the previous section, there are many covariant
clock time operators. For uniqueness additional,
physically motivated conditions are needed. Requiring minimal variance by itself does not make $\hat{T}$ unique, not even in the case of non-degenerate
spectrum of $\hat{H}$, since in general it may not be possible
to fulfill this requirement simultaneously for all states
with second moment unless, in addition,
one restricts the set of functions $b_i$ by symmetry
requirements, as we shall now discuss.

The time reversal operator,  here denoted by $\hat\Theta$, is an 
anti-unitary operator. If the dynamics is time reversal invariant, it
is natural to demand
\begin{equation} \label{eq3.1}
\hat\Theta\, \hat{T}\, \hat\Theta = - \hat{T},
\end{equation}
and similarly for the probability density. By Eq. (\ref{eq1.7bc}) this
implies 
\begin{equation} \label{eq3.2}
\hat\Theta \,\hat{\Pi}_0\hat \Theta = \hat{\Pi}_0.
\end{equation}

It will now be shown for the non-degenerate eigenvalue case that time
reversal invariance of the Hamiltonian $\hat H$ and of $\hat\Pi_0$, and minimal 
$\Delta T$ together imply uniqueness of $\hat{T}$ and $\hat\Pi_t$. For
each eigenvalue $E$ of $\hat H$ one can choose a $\hat\Theta$ invariant
eigenvector, denoted by $ |E_\Theta \rangle$,
\begin{equation} \label{eq3.3}
\hat\Theta |E_\Theta  \rangle = |E_\Theta \rangle.
\end{equation}
This means a specific choice of phase factor and a real function
in position space. Eq. (\ref{eq3.2})
implies $\hat{\Pi}_0 = 1/2 (\hat{\Pi}_0 + \hat\Theta\,
\hat{\Pi}_0\,\hat \Theta)$, and the general 
form of $\hat{\Pi}_0$ in Eq. (\ref{eq2.1}) then implies that $b_i(E)$ can be
chosen real. Then, from Eqs. (\ref{eq2.3a}) and (\ref{eq2.4}), one finds
\begin{equation} \label{eq3.3a}
 \langle \psi |\hat{T}| \psi \rangle = \int dE\,  \overline{\psi(E)}\,
\frac{\hbar}{i}\, \psi^\prime (E),
\end{equation}
\beqa \label{eq3.4}
\nonumber
&\mbox{second moment}
 = \hbar^2\int dE\, |\psi^\prime (E)|^2
\\
& +
\hbar^2\sum_i \int dE\, |\psi|^2\, |b_i^\prime(E)|^2.
\eeqa
Hence $\Delta T$ minimal means in this case that the second moment is
minimal, and the latter holds if and only if $b_i^\prime (E) \equiv 0$, i.e.
$$
b_i(E) \equiv c_i,~~~\sum ~c_i^2 = 1,
$$
by Eq. (\ref{eq2.3}). Inserting this into Eq. (\ref{eq2.1}) one sees
that the functions $b_i$  
can be replaced by the single function $b(E) \equiv 1$. Thus one
obtains
\begin{equation} \label{eq3.5}
\begin{split}
\hat{\Pi}_0& =\frac{1}{2\pi\hbar}\int dE~dE^\prime \,|E_\Theta \rangle
\langle E_\Theta ^\prime|,
\\ 
\hat\Pi_t &= \frac{1}{2\pi\hbar}\int dE\,dE'\,
e^{-i(E-E')t/\hbar}|E_\Theta \ra\la E_\Theta '|,
\\
\hat T &= \int dt\, \hat\Pi_t,
\end{split}
\end{equation}
with time reflection invariant $|E_\Theta \rangle$. 
The (non-orthogonal) eigenfunctions, 
$|\tau\ra$, of  $\hat T$ with eigenvalue $\tau$ are given by
\beq
\label{t0}
|\tau\ra=\frac{1}{\sqrt{2\pi\hbar}}\int_0^\infty dE e^{-i E
  \tau/\hbar}|E_\Theta\ra, 
\eeq
and $\hat T$  can be written as
\beq
\label{ta0}
\hat{T}=\int_{-\infty}^{\infty} d\tau \, \tau|\tau\ra\la \tau|.
\eeq
Therefore uniqueness holds
in the non-degenerate case if time-reversal invariance 
and minimal $\Delta T$ are demanded. 

In the degenerate eigenvalue case
this is no longer true and one needs additional conditions to obtain
uniqueness, as discussed elsewhere \cite{Gerhard}. 
Here we simply state some results.  With a reflection
invariant potential in one dimension, the clock time operator becomes
unique and can be explicitly determined if, in addition to covariance
under time reversal and minimal variance, one also demands invariance
under space reflection. With a rotation invariant potential in three
dimension, the time operator becomes unique and can be 
explicitly determined if, in addition to covariance under time
reversal and minimal variance, one also demands invariance under
rotations and reflection $x_1 \to -x_1$. Analogous results hold for
arrival time operators. In particular,  a
generalization of the result of Ref. \cite{Kij} is obtained \cite{Gerhard}. 
\section{Application to arrival times}\label{arrival}
Evidently the techniques of the previous sections can be applied in a
completely analogous way to the study of arrival-time operators. 
To illustrate this we consider in the following the motion of a particle on the
half-line $x\geq 0$, without additional potential, and  study
its arrival times at the origin and  at an arbitrary point.  

In the classical case an incoming free particle  of energy $E$ is
reflected at 
the origin and then travels back to infinity. Hence, for each point
$a\neq 0$, there is a first and second time of arrival which we denote by
$t^a_1$ and $t^a_2$. For the time reversed trajectory the first
arrival at $a$ is at time $t^{ a}_{\theta,1} = -t^a_2$ and the second
arrival at time $t^{ a}_{\theta,2} = -t^a_1$, as is easily
calculated. For the origin, $a=0$, there is only one arrival and 
\beq \label{reversed}
t^{0}_{\theta} = -t^0.
\eeq

The corresponding
arrival-time operator for arrivals at the origin is denoted by $\hat
T^{A}_f$. It is natural to demand the analogous relation to
Eq. (\ref{reversed}), i.e.   
\beq \label{qreversed}
\hat \Theta \hat T^{A}_f\hat \Theta = -\hat T^{A}_f,
\eeq
and  time reversal invariance of
$\hat\Pi^A_{f,0}$, where $\hat\Pi^A_{f,t}$ is the associated
probability density operator.

If $a\neq 0$ a classical free particle on the positive half-line,
coming in from  
infinity with velocity $|v|$, arrives first at time $t_1^{\,a}$ at the
point $a$, and then at time $t^{\, 0}$ at the origin,  
\beq \label{heuristic1}
t_1^{\,a} = t^{\,0} - a/|v|.
\eeq
If $\hat T^{A}_{1}$ denotes the corresponding time
operator for the first arrival at $a$ we may demand
\beq \label{heuristic2}
\hat T^{A}_{1} = \hat T^{A}_f -a/|\hat v|,
\eeq
where $|\hat v| = \sqrt{2\hat H/m}$ is the velocity operator.
\subsection {Free particle on a half-line}
We first consider arrivals at the origin for free motion on
the half-line $x\geq 0$, with reflecting boundary 
conditions at $x=0$. The eigenfunctions  can be labeled by the energy 
$E=k^2\hbar^2/(2m)$.  Real, and thus $\hat \Theta$ invariant,
eigenfunctions for energy $E$ which vanish at the origin are
\beq \label{eq4.1aa}
\la r|E_f\ra
=\frac{i}{\hbar}\sqrt{\frac{m}{2\pi k}}(e^{-ikr}-e^{ikr}), 
\eeq
where the
subscript $f$ in $|E_f\ra$ refers to the free Hamiltonian and where we
have written $r$ to indicate $r\equiv x\geq 0$. These  
eigenfunctions are normalized as $\la E_f|E_f'\ra=\delta(E-E')$ on the
half-line.

For the probability density  operator for arrivals
at the origin  invariance under time reversal means 
\beq \label{origin}
\hat\Theta \,\hat{\Pi}^{A}_{f,0}\,\hat\Theta = \hat{\Pi}^{A}_{f,0}.
\eeq
By the results of the last section, the  operators
$\hat{\Pi}^{A}_{f,t}$ and  $\hat T^{A}_f$ 
become unique if  invariance under time reversal holds
and minimal variance is assumed. From Eq. (\ref{eq3.5})
one obtains, with a change $t \to -t$ and replacing $|E_\Theta \ra$ by $|E_f\ra$,
\beqa \label{eq4.1a}
\hat\Pi^{A}_{f,\,t} &=& \frac{1}{2\pi\hbar}\int dE\,dE'\,
e^{i(E-E')t/\hbar}|E_f\ra\la E'_f|,
\\
\hat T^{A}_f &=& \int dt\,t\, \hat\Pi^{\,0}_{f,\,t}.\nonumber
\eeqa
This arrival time operator is just the negative of the clock
time operator of Eq. (\ref{eq3.5}), with Eqs.~(\ref{t0}) and
(\ref{ta0}) holding correspondingly.

Note that the vanishing of the wave function at $r=0$ is not an obstacle
to define these operators in a physically meaningful manner. 
A similar situation is found for antisymmetrical wavefunctions on the 
full line. It was shown in \cite{HSMN} that the ideal time-of-arrival distribution follows in a limiting process from an operational measurement model 
that considers explicitly a weak and narrow detector.    

We now turn to first arrivals at $a\neq 0$. Using Eq. \ref{CCR}, a simple calculation shows that
\beq
\label{shift}
e^{iam|\hat v|/\hbar}\,\hat{T}^{A}_f\, e^{-iam|\hat v|/\hbar} =
\hat{T}^{A}_f - a/|\hat v|.
\eeq
Since the right-hand side equals $\hat T^A_{1}$, by
Eq. (\ref{heuristic2}), this implies an analogous relation for the
probability density operator, $\hat{\Pi}^{A}_{1,t}$, for $\hat T^A_{1}$,
\beq
\label{shift2}
\hat{\Pi}^{A}_{1,t} = e^{iam|\hat v|/\hbar}\,\hat{\Pi}^{A}_{f,t}\,
e^{-iam|\hat v|/\hbar}.
\eeq
Using Eq. (\ref{eq4.1a}) this can be written as
\beq
\label{shift3}
\hat{\Pi}^{A}_{1,t} = \frac{1}{2\pi\hbar}\int dE\,dE'\,
e^{i(E-E')t/\hbar} e^{i(k - k')a}|E_f\ra\la E'_f|,
\eeq
which explicitly gives the temporal probability density operator for
the first arrival at the point $a$ of a free particle on the positive
half-line. For $a \to 0$ one recovers Eq.~(\ref{eq4.1a}).  
\subsection{Asymptotic states and Smith's delay time}
We now apply the free-particle result in Eq. (\ref{eq4.1a}) to the
asymptotic states  of a particle in a potential on the
half-line whose Hamiltonian  has no bound states and to which
scattering theory applies. Although for fixed $E$ the eigenstate is
unique up to a 
phase, there are physically relevant eigenstates  $|E_\pm\ra$  which
correspond to an incoming (+) and outgoing (-) plane wave,
respectively, as well as the $\Theta$ invariant state, denoted by
$|E_\Theta\ra$. Their relation and asymptotics are $|E_-\ra
=\hat\Theta |E_+\ra$ and, with the scattering phase shift $\delta=\delta(E)$, 
\beqa
\la r|E_+\ra&\sim& \frac{1}{\hbar}\sqrt{\frac{2m}{k\pi}}\frac{i}{2}(e^{-ikr}-e^{2i\delta}e^{ikr}), 
\nonumber\\
\la r|E_-\ra&=&\overline{\la r|E_+\ra} = e^{-2i\delta}\la r|E_+\ra,
\nonumber\\
\la r|E_\Theta\ra&=&e^{-i\delta}\la r|E_+\ra.
\label{eq4.1ab}
\eeqa
The  M{\o}ller operators $\hat \Omega_\pm$ satisfy
\beqa \label{eq4.3a}
\hat{\Omega}_\pm &\equiv& \lim_{t\to\mp\infty}e^{i\hat{H}t/\hbar} 
e^{-i\hat{H}_ft/\hbar}=\int_0^\infty dE\,|E_\pm\ra\la E_f|,
\nonumber\\
|E_\pm\ra &=&\hat \Omega_\pm |E_f\ra.
\eeqa
The freely moving asymptotic states $|\psi_{in}\ra$ and $|\psi_{out}\ra$ are
mapped by $\Omega_\pm$  to the actual state $|\psi\ra$, 
\beqa \label{eq4.6a}
|\psi\ra&=&\hat{\Omega}_\pm\,|\psi_{\stackrel{in}{out}} \ra
\\
|\psi_{out}\ra&=&\hat{S}\,|\psi_{in}\ra \nonumber
\eeqa
where $\hat{S}=\hat{\Omega}_-^\dagger\hat{\Omega}_+$ is the $S$
operator. Note that, by Eq.~(\ref{eq4.1ab}),
\beq 
\hat{S}=\int_0^\infty dE\, |E_f\ra e^{2i\delta} \la E_f|,
\eeq
so that
$e^{2i\delta}$ is the eigenvalue of $\hat{S}$ for the state $|E_f\ra$.

It is  convenient to  introduce also the operator
\beq \label{eq4.5a}
\hat \Omega_\Theta \equiv\int_0^\infty dE |E_\Theta\ra\la E_f|
\eeq
and define operators $\hat{T}^{\, A}_{\pm,\Theta}$ by 
\beqa \label{eq4.7a}
\hat{T}^{\, A}_{\pm,\Theta} &\equiv&
\hat{\Omega}_{\pm,\Theta}\,\hat{T}^{\,
  0}_f\,\hat{\Omega}_{\pm,\Theta}^\dagger\\ 
&=& \int dt\, t\int dE\,dE'\, e^{i(E-E')t/\hbar}|E_{\pm,\Theta}\ra\la
E'_{\pm,\Theta}| \nonumber.
\eeqa
The last line shows that $-\hat{T}^{\, A}_{\pm,\Theta}$ are possible clock time
operators for the particle in the potential.
Since the states $|E_{\pm,\Theta}\ra$ differ only by a phase, the same
calculation that leads to Eq. (\ref{eq1.16}) gives
\beq \label{eq4.2aa} 
\hat T^{\, A}_{\pm} =\hat{T}^{\, A}_{\Theta} \mp \hbar\int dE\, 
\frac{\partial\delta}{\partial E}\,|E_\Theta\ra\la E_\Theta|.
\eeq
{}From Eq. (\ref{eq4.6a}) it follows that the expectation values of
$\hat{T}^{\, A}_{+}$, $\hat{T}^{\, A}_{-}$ and $\hat{T}^{\,
A}_{\Theta}$ may be interpreted in terms of the asymptotic  
states and the free-motion arrival-time operator $\hat{T}^{\, A}_f$,  
\beq
\la\psi|\hat{T}^{\, A}_{+,-,\Theta}|\psi\ra =\la
\psi_{in,out,io}|\hat{T}^{\,
  A}_f|\psi_{in,out,io}\ra,     
\label{last}
\eeq
where the freely moving state $|\psi_{io}\ra$ is defined by
\beq \label{interpolation}
|\psi_{io}\ra \equiv\hat S^{1/2}|\psi_{in}\ra, 
\eeq
and can be considered as an interpolation between 
$|\psi_{in}\ra$ and $|\psi_{out}\ra = \hat S |\psi_{in}\ra$. 
With Eq. (\ref{eq4.5a})
one can write
\beq \label{eq4.5aa}
|\psi_{io}\ra = \Omega_\Theta^\dagger |\psi\ra.
\eeq
Taking expectation values of Eq. (\ref{eq4.2aa}) with $|\psi\ra$ and
using Eqs. (\ref{last}) and (\ref{eq4.3a}),
together with the fact that $|E_{\pm,\Theta}\ra\la E_{\pm,\Theta}| $
all coincide since the phases drop out, yields
\beq \label{4.9a}
\begin{split}
\la\psi_{\stackrel{in}{out}}|\hat{T}^{\, A}_f|\psi_{\stackrel{in}{out}}\ra = 
\la\psi_{io}|
\,\hat{T}^{\, A}_{\, f}\,|\psi_{io}\ra
\mp \hbar\int
dE\,\frac{\partial\delta}{\partial E}\,\big|\la E_f| \psi_{in}\ra\big|^2.
\end{split}
\eeq
One sees from this that the mean arrival time for the interpolating state
$|\psi_{io}\ra$ lies between those of the ingoing and outgoing wave. 
From Eq.~(\ref{4.9a}), 
\beq \label{4.10a}
\la
\psi_{out}|\hat{T}^{\, A}_f|\psi_{out}\ra  -
\la\psi_{in}|\hat{T}^{\, A}_f|\psi_{in}\ra = 2 \hbar\int
dE\,\frac{\partial\delta}{\partial E}\,\big|\la E_f| \psi_{in}\ra\big|^2.
\eeq
The right-hand side of the last equation is the scattering time delay
of Smith \cite{Smith} and it shows that the time for the outgoing wave is
shifted with respect to the time for the ingoing wave by the
scattering time delay. An example is shown in Figs. 1 and 2.  

{\it Time-reversal:} The behavior of $T^{\, A}_\pm$ with respect to 
time-reversal is determined by acting with the anti-linear operator 
$\hat{\Theta}$,  
\beq
\hat{\Theta}\, \hat{T}^{\, A}_\pm\,\hat{\Theta} =-\hat{T}^{\, A}_\mp, 
\label{mm}
\eeq
whereas $\hat{T}^{\, A}_{\Theta}$ simply changes sign.
The operators $\hat{T}^{\, A}_\pm$ do not simply change sign under
time-reversal  
as $\hat{T}^{\, 0}_{\, \Theta}$ does, but their behavior in Eq. (\ref{mm}) (changing
the sign and  exchanging the operators) is perfectly physical:  
the time reversal of a trajectory which moves towards the origin 
is a trajectory in the same location but moving away from the origin.  
If the original incoming trajectory requires a certain time $\tau$ to arrive 
at the origin (with free motion),  
the reversed trajectory is outgoing, and departed from the origin 
at $-\tau$. These operators provide in summary information of the
free-motion dynamics of incoming and outgoing asymptotes of the state, 
and scattering time delays \cite{Leon,toapot}. Thus, although the operator
$\hat{T}^{\, A}_{\, \Theta}$ is unique when one applies the criteria of the 
previous section  it does not supersede  $\hat{T}^{\, A}_{\pm}$ since
it does not describe  
the same physics, and all three operators have their own legitimacy.         
\begin{figure}[t]
\vspace{-4cm}\hspace{-1cm}
\includegraphics[height=13.cm,angle=0]{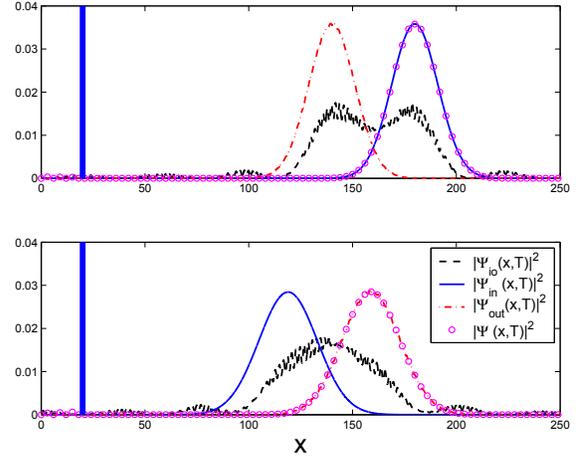}
\vspace{-3cm}
\caption{\label{f1} 
(Color online) Probability densities before ($t=0$) and after the collision
($t=190$) 
with a delta barrier. Dimensionless units with $m=\hbar=1$.  
The initial wave packet is $\psi(k)=N[1-\exp(-\beta k^2)]\exp[-(k-k_0)^2/(4\Delta_k^2)]
\exp(-ikx_0)\theta(k)$, where $N$ is the normalization constant
and $\theta$ (here) the Heaviside step function; initial wavenumber $k_0=-\pi/2$,
$\Delta_k=0.045$, $\beta=1/2$; $V=20\delta(x-20)$; initial center of the wave packet
$x_0=180$. The delta potential is rather opaque so the the outgoing packet is advanced with respect to the incoming state.} 
\end{figure}
\begin{figure}[t]
\vspace*{-3cm}\hspace{-1cm}
\includegraphics[height=11.cm,angle=0]{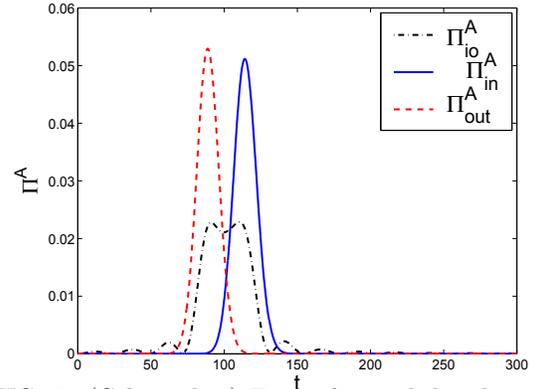}
\vspace{-3cm}
\caption{\label{f2} 
(Color online) Time of arrival distributions for arrivals at $x=0$ 
corresponding to the previous figure.} 
\end{figure}
\section{Application to Lyapunov operators
in Quantum Mechanics}\label{Lyapunov}
In Ref. \cite{Strauss} an  operator $\hat L$ was called  a
Lyapunov operator if for any 
normalized $| \psi \rangle$ and $| \psi_t \rangle \equiv
e^{-i\hat{H}t/\hbar}| \psi \rangle$, the expectation value
$\langle \psi_t |\hat L|\psi_t \rangle$ is monotonically
decreasing  to 0 as $t \to \infty$ and goes to
1 for $t \to -  \infty$. Ref. \cite{Strauss} considered the case of a
Hamiltonian $\hat H$ with purely 
continuous eigenvalues ranging from 0 to infinity and degeneracy
parameter $j$. The particular Lyapunov operator suggested there can be written
as
\be \label{eq6.1}
\hat{L}_S = \frac{i}{2\pi\hbar} \sum_j \int_0^\infty dE
\int_0^\infty dE^\prime \frac{|E, j \rangle \langle E^\prime, j|}{E-
  E^\prime + i \varepsilon}.
\ee
More generally, one may call a bounded operator $\hat L$ a Lyapunov
operator if $\langle \psi_t |\hat L|\psi_t \rangle$ is just monotonically
decreasing, without specifying limits. However, it will be shown below,
after Eq. (\ref{eq6.2}), that, without loss 
of generality, one can always assume the above limiting behavior from
1 to 0 as $t$ goes from $ - \infty$ to $ + \infty$.

The above notion does not quite correspond to Lyapunov functionals  used in
Ref. \cite{Sewell} to define irreversibility and an 
arrow of time, since time reversal invariance of the functional was
assumed there in order to have neutrality with respect to past and
future. It will be shown further below that there are no time reversal
invariant Lyapunov operators if the Hamiltonian is time reversal
invariant.

It is clear that the above properties do not define $\hat{L}$ in
Eq. (\ref{eq6.1}) 
uniquely. For example, one can introduce phases  and
still get a Lyapunov operator. In this section we are going to
determine the most general form of $\hat{L}$ for a Hamiltonian
$\hat{H}$ with a purely (absolutely) continuous spectrum and give
conditions under which it becomes unique. It will also be seen that
to each $\hat{L}$ there is an associated covariant time
operator $\hat{T}_L$.

To show that one can assume the above limit behavior we put, for a
given general Lyapunov operator $\hat{L}$, 
\be \label{eq6.2}
\hat{L}_t \equiv e^{-i \hat{H}t/\hbar} \hat{L} e^{ i \hat{H}t/\hbar}
\ee
so that $\hat{L}_t$ is monotonically increasing, by the
monotonic decrease of $\langle \psi_t |\hat{L} |\psi_t \rangle$. From
the boundedness of $\hat{L}$ and from monotonicity it follows that
$\hat{L}_{\pm \infty}$ exist as operator limits in the weak sense,
i.e. for expectation values. Moreover, $\hat{L}_{\pm \infty}$ commutes
with $e^{-i \hat{H}t/\hbar}$, and therefore $\hat{L}'\equiv
\hat{L}-\hat{L}_{- \infty}$ is also a  Lyapunov operator, with
$\hat{L}'_t\geq 0$. Then $\hat{L}''\equiv \hat{L}'^{-1/2}
\hat{L}'\hat{L}'^{-1/2}$ is a Lyapunov operator satisfying $\hat{L}''_{- \infty}=0$
and $\hat{L}''_{ \infty}=1 $ so that $\langle \psi_t |\hat L''|\psi_t
\rangle$ is monotonically decreasing  from 1 to 0, proving the above claim.

To determine the general form of $\hat{L}$ with such a limit behavior
for $t \to \pm \infty $, we note that by monotonicity 
\be \label{eq6.3}
\hat{\Pi}^L_t \equiv \frac{d}{dt}{\hat{L}}_t = e^{-i \hat{H}t/\hbar}
\frac{-i}{\hbar} [\hat{H}, 
\hat{L}]e^{i \hat{H}t/\hbar} \ge 0,
\ee
i.e. expectation values of $\dot{\hat{L}}_t$ are non-negative for
all $t$, in particular 
\be \label{eq6.4}
\hat{\Pi}^L_0 = \frac{-i}{\hbar} [\hat{H}, L] \ge 0
\ee
where the commutator is again to be understood in the weak sense via
matrix elements and where $\hat{\Pi}^L_0$ is in
general not an operator but only a bilinear form, as in Eq. (\ref{eq1.7bc}).
{}From Eq. (\ref{eq6.3}) and from $\hat{L}_{- \infty} = 0$
one obtains
\be \label{eq6.5}
\hat{L}= \int_{- \infty}^0 {dt}\, e^{-i \hat{H}t/\hbar}\, \hat{\Pi}^L_0\,
e^{i \hat{H}t/\hbar}.
\ee
{}From Eq. (\ref{eq6.3}) one sees that
\be \label{eq6.6}
\Pi_L (t; \psi) \equiv \langle \psi |\hat{\Pi}^L_t |\psi \rangle\, \ge\, 0
\ee
is a non-negative density which integrates to 1 for each normed state,
i.e. it can be regarded as a probability density and
hence $\hat{L}_t$ behaves like the cumulative probability operator
$\hat{F}_\tau$ in Eq. ({\ref{eq1.1}}). Therefore,
\be \label{eq6.7}
\hat{T}_L \equiv \int dt\, t\, e^{-i \hat{H}t/\hbar}\,
\hat{\Pi}^L_0\, e^{i \hat{H}t/\hbar}
\ee
is an analog of the time operator $\hat{T}$ in Eq. (\ref{eq1.4}).
Alternatively, $1- \hat L_{-t}$ behaves as the cumulative arrival
probability operator $\hat F^A_t$ in Eq. (\ref{eq5.7a}).
\\[.3cm]
{\bf Example:} Let $\hat{\Pi}^L_0$ given by Eq. (\ref{eq1.14}). Then,
by Eq. (\ref{eq6.5}), $\hat{L}$ is given by
\be \label{eq6.8}
\hat{L} = \frac{1}{2\pi\hbar}\int^0_{- \infty} dt \int dE~dE^\prime e^{-i(E-
  E^\prime)t/\hbar} |E \rangle \langle E^\prime|,
\ee
which is readily seen to agree with $\hat{L}_S$ in
Eq. (\ref{eq6.1}) in the case of non-degeneracy.

{}For free motion on the half-line, with $|E\ra=|E_f\ra$ from
Eq.~(\ref{eq4.1aa}), the Lyapunov
property of this example simply reflects the
monotonous accumulation of arrivals at the origin since
 a change of integration variable gives
\beq
\label{accum}
\la \psi_t|1 - \hat{L}|\psi_t\ra=\int_{-\infty}^t dt'\, \la
\psi|\hat\Pi^0_{f,\,t'}|\psi\ra.  
\eeq 
With a potential on the half-line and taking $|E\ra =|E_\pm\ra$ of the
previous section one obtains the accumulation of arrivals of  the freely
moving packets
$|\psi_{in}\ra$ and $|\psi_{out}\ra$, and for $|E\ra =|E_\Theta\ra$
the corresponding
accumulation of arrivals for $|\psi_{io}\ra$.    


The most general form of $\hat{L}$ is obtained from the most general
form of $\hat{\Pi}^L_0$ which is given by Eqs. (\ref{eq2.8}) and
(\ref{eq2.9}). If $\hat{\Pi}^L_0$ is known then $\hat{L}$ is given by
Eq. (\ref{eq6.5}), and in this way one obtains the most general form
of the Lyapunov operator $\hat L$ with the above limit behavior for
$t\to \pm\infty$. 
Uniqueness of $\hat{L}$ may be achieved for particular Hamiltonians by
demanding, e.g. time reflection invariance of $\hat T_l$, special symmetries
and minimal variance $\Delta T_L$, as in Sections \ref{uniqueness} and
\ref{arrival}. 

We finally show that for a time reversal invariant Hamiltonian  there is
no nontrivial time reversal invariant Lyapunov operator. Indeed, if
$\hat\Theta\,\hat H \,\hat\Theta = \hat H$ and $\hat\Theta\,\hat L
\,\hat\Theta =\hat L$ then one obtains, for initial state
$\hat\Theta\,|\psi\ra \equiv |(\hat\Theta\psi)\ra$,
\beq \label{eq6.9}
\la(\hat\Theta\psi)_t|\hat L|(\hat\Theta \psi)_t\ra =
\la\psi_{-t}|\hat L| \psi_{-t}\ra~  
\eeq 
by the anti-unitarity of $\hat\Theta$. Now, for increasing $t$, the
expression on the left-hand side
decreases while the one on the right-hand side increases. This is
only possible if both sides are constant in $t$. Alternatively, one
can conclude from Eq. (\ref{eq6.4}) that both $\hat{\Pi}^L_0$ and
$\hat\Theta\,\hat{\Pi}^L_0\,\hat\Theta = -\hat{\Pi}^L_0$ are positive
operators, which is only possible if $\hat{\Pi}^L_0=0$. This means
that $\hat L$ commutes with $\hat H$, which also leads to the
constancy of both sides in Eq. (\ref{eq6.9}).
\section{Discussion and outlook}
We have provided the most general form of covariant,
normalized time operators. This is important to set 
a flexible framework where physically motivated conditions
on the observable may be imposed. The application examples 
include clock time operators, time-of-arrival operators 
and Lyapunov operators.

Experimentally, a number of interesting open questions remain for
quantum clocks  and arrival-time measurements.  
For example, quantum clocks are basically  
quantum systems with an observable that evolves linearly with time. 
To evaluate the possibility to compete with current atomic clocks
\cite{MLM}, 
the observable must be realized in a specific system.   
We have described an ideal observable (by imposing 
antisymmetry with respect to time reversal and minimal variance)
and the analysis of the operational realization is now pending. 
A similar analysis for the ideal arrival time-of-arrival distribution 
of Kijowski has been carried out in terms of an operational 
quantum-optical realization with cold atoms (cf. Ref. \cite{toatqm2}
for a review). Indeed, cold atoms and quantum optics   
offer examples of times of events (other than arrivals), such as 
jump times, excitation times, 
escape times, admitting 
a treatment in terms of covariant observables. Modeling and understanding these quantities and their statistics 
may improve our ability to manipulate or optimize dynamical processes.   

On the theory side, an open question is how to adapt the proposed framework,
possibly in combination with previous investigations
\cite{crossstates,Leon,toapot,HSMN,Galapon06,Galapon08,Galapon}, to 
arrival times when a particle moves in a potential. 

Finally, we have shown that Lyapunov operators follow naturally from
covariant time observables. Associated to time-of-arrival operators, they
account for the monotonous accumulation of arrivals for freely moving 
asymptotic states from the infinite past independently of the state chosen.
Note that the ``infinite past'' here is an idealized construct since
it must be assumed that the wave has been evolving forever, ignoring
the fact that in practice  
the state may have been prepared at some specific instant. In other
words the Lyapunov operator does not depend on that preparation
instant, and when applied to the state it takes into account its
idealized (not necessarily actual) past, 
whether or not that past has been fully or partially realized.         

We have also shown at the end of the last section that in theories
with a time reversal invariant Hamiltonian there are no time reversal
invariant Lyapunov operators. In Ref. \cite{Sewell} it was argued
that in order to characterize a system as irreversible and single out
a direction of time a Lyapunov functional should be time reversal invariant. 
Hence, if one accepts this view of Ref. \cite{Sewell} 
then, by our result, quantum mechanics for finitely many particles 
should indeed not be irreversible and should not exhibit an arrow of
time if the Hamiltonian is time reversal invariant.
\section*{Acknowledgments}
We thank L. S. Schulman and J. M. Hall for discussions. 
We also acknowledge the kind hospitality of the Max Planck 
Institute for Complex Systems in Dresden, and   
funding by the Basque Country University UPV-EHU (GIU07/40), and the  
Ministerio de Educaci\'on y Ciencia Spain (FIS2009-12773-C02-01). 

\appendix
\section{Minimal variance and non-uniqueness of time operator}\label{B}
We show  for the case of a non-degenerate spectrum of $\hat{H}$  that
minimal variance alone does not imply uniqueness of $\hat T$. 
We first consider a state $|\psi\rangle$ such that, with a given
choice of generalized eigenvectors,  $\langle E |\psi \rangle
\equiv \psi (E)$ is real. Then the first term on the r.h.s of
Eq. (\ref{eq2.3a}) is the integral of a total derivative and therefore
vanishes, as does the third term on the r.h.s. of
Eq. (\ref{eq2.4}), by Eq (\ref{eq2.3}). Thus
\beqa \label{A.20}
\Delta T^2 &=& \int dE \, |\psi^\prime|^2 + \sum_i \int dE \,  |b_i^\prime|^2
\,|\psi|^2
\nonumber\\
&-&\left(\int dE\, |\psi|^2\, i \sum_i b_i \overline{b_i^\prime}\right)^2.
\eeqa
By Schwarz's inequality the last term can be estimated as
\beqa \label{eqA.21}
&&\Big| \sum_i \int dE \,|\psi|^2\, b_i\, \overline{b_i^\prime}\,
\Big|^2 
\nonumber\\
&&\leq \sum_i \int dE\, |\psi|^2\, |b_i|^2 \cdot \sum_i \int dE\,
|\psi|^2\, |b_i^\prime|^2, 
\eeqa
where the first sum on the r.h.s. yields 1 and the equality sign holds
if and only if
\be \label{eqA.22}
b_i^\prime (E) = \gamma\, b_i (E)~~,~~ \gamma = \mbox{constant},
\ee
which implies
\be \label{eqA.23}
\sum b_i\,\overline{b_i^\prime} = \bar{\gamma} \sum b_i \,\overline{b_i}
= \bar{\gamma}. 
\ee
Since the l. h. s. is purely imaginary, from Eq. (\ref{eq2.3}), this
implies $\gamma = i \lambda$ with $\lambda$ real. Thus, for real
$\psi(E)$, 
\be \label{eqA.24}
\Delta T^2 \geq \int dE |\psi^\prime (E)|^2,
\ee
with equality holding if and only if Eq. (\ref{eqA.22}) holds with
$\gamma = i \lambda~,~\lambda$ real, i.e. if and only if
\begin{eqnarray} \label{eqA.25}
b_i (E) &=& c_i\, e^{i \lambda E}~,~~~~\lambda~\mbox{real} \\ \nonumber
\sum_i b_i (E)\, \overline{b_i(E)} &=& \sum |c_i|^2 = 1.
\end{eqnarray}
These functions give the same time operator and density as the single
function
\be \label{eqA.26}
b(E) = e^{i \lambda E}.
\ee
With this choice $\Delta T^2$ becomes minimal for {\it real}
$\psi(E)$.  

For a state given by $e^{i \varphi(E)} \psi(E)$, with real $\psi(E)$ and
$\varphi(E)$, the same argument gives, upon replacing $b_i$ by $e^{-i
\varphi(E)}\,b_i$, that one has minimal variance if and only if
\be \label{eqA.27}
b_i (E) = c_i\, e^{i(\lambda - \varphi(E))}.
\ee
This differs from Eq. (\ref{eqA.25}), as does the analog $e^{i(\lambda
  - \varphi(E))}$ of the single function in Eq. (\ref{eqA.26}). 

Hence among the set of all allowed functions $b_i(E)$ there is no
choice of functions such that $\Delta T$ becomes minimal for {\em all}
states with finite second moment.


\begin{thebibliography}{999}     
%
\bibitem{b1} J. G. Muga, R. Sala Mayato and I. L. Egusquiza (eds.),
{\it Time in Quantum Mechanics}, Vol. 1, Lect.
Notes Phys. \textbf{734}, Springer-Verlag, Berlin  2008.  
\bibitem{b2} J. G. Muga, A. Ruschhaupt and A. del Campo, {\it Time in
  Quantum Mechanics}, Vol. 2,  Lect. Notes Phys. \textbf{789},
  Springer-Verlag, Berlin 2009.  
\bibitem{Holevo} A. S. Holevo, {\it Probabilistic and statistical aspects
of quantum theory}, North Holland, Amsterdam, 1982.    
\bibitem{Hall} M. J. W. Hall, eprint arXiv:0802.2682.
\bibitem{dwell} J. Mu\~noz, I. L. Egusquiza, A. del Campo, D. Seidel
and J. G. Muga, Lect. Not. Phys. \textbf{789}, 97 (2009). 
\bibitem{clocks} R. Sala Mayato, D. Alonso, and I. L. Egusquiza,
Lect. Notes Phys. \textbf{734}, 235 (2008). 
\bibitem{Smith} F. T. Smith, Phys. Rev. \textbf{118}, 349 (1960). 
\bibitem{Strauss} Y. Strauss, J. Silman, S. Machnes, L.P. Horwitz,
  eprint arXiv:0802.2448. 
\bibitem{Sewell} G. L. Sewell, {\it Quantum Mechanics and its Emergent
  Macrophysics}, Princeton University Press, Princeton and Oxford,
2002, p. 84.
\bibitem{Hall08} M. J. Hall, J. Phys. A: Math. Theor. \textbf{41},
  255301 (2008). 
\bibitem{Caves} S. L. Braunstein, C. M. Caves and G. J. Milburn, 
Ann. Phys. (NY) \textbf{247}, 135 (1996). 
\bibitem{Hall95} M. J. Hall in ``Quantum Communications and
  Measurement'',  ed. by V. P. Belavkin,  
O. Hirota and R. L. Hudson (New
York: Plenum, New York 1995) p. 53. 
\bibitem{MugaLeavens} J. G. Muga and  C. R. Leavens,
  Phys. Rep. \textbf{338}, 353 (2000). 
\bibitem{toatqm2} A. Ruschhaupt, J. G. Muga and G. C. Hegerfeldt, 
Lect. Not. Phys. \textbf{789}, 65 (2009). 
\bibitem{Werner} R. Werner, J. Math. Phys. \textbf{27}, 793 (1986).
\bibitem{Kij} J. Kijowski, Rep. Math. Phys. \textbf{6}, 362 (1974).
\bibitem{Gerhard} G. C. Hegerfeldt and J. G. Muga, in preparation
\bibitem{HSMN} G. C. Hegerfeldt, D. Seidel, J. G. Muga, and B. Navarro, 
Phys. Rev. A \textbf{70}, 012110 (2004).
\bibitem{Leon} J. Le\'on, J. Julve, P. Pitanga, and F. J. de Urr\'\i es,
Phys. Rev. A \textbf{61}, 062101 (2000).
\bibitem{toapot} A. D. Baute, I. L. Egusquiza and J. G. Muga, Phys. Rev. A 
\textbf{64}, 012501 (2001). 
\bibitem{MLM} J. Mu\~noz, I. Lizuain and J. G. Muga, 
Phys. Rev. A \textbf{80}, 022116 (2009). 
\bibitem{crossstates} A. D. Baute, R. Sala Mayato, J. P. Palao, 
J. G. Muga and I. L. Egusquiza, Phys. Rev. A \textbf{61}, 022118 (2000). 
\bibitem{Galapon06} E. A. Galapon, Int. J. Mod. Phys. A \textbf{21}, 6351 (2006).
\bibitem{Galapon08} E. A. Galapon and A. Villanueva, J. Phys. A:Math. Theor. \textbf{41}, 455302 (2008).
\bibitem{Galapon} E. A. Galapon, Lect. Notes Phys. \textbf{789}, 25
(2009), and references therein. 
%
\end{thebibliography}
\end{document}